**Geometry-Driven Polarization Control in Ferroelectric Nematic Liquid Crystals**

*Kazuma Nakajima*, Hirokazu Kamifuji, Hirotsugu Kikuchi, Kenjiro Fukuda* and Masanori Ozaki*

Kazuma Nakajima, Hirokazu Kamifuji, Kenjiro Fukuda and Masanori Ozaki
Division of Electrical, Electronic, Informational Engineering, Graduate School of Engineering, The University of Osaka, 2-1 Yamada-oka, Suita, Osaka 565-0871, Japan
E-mail: nakajima.kazuma@eei.eng.osaka-u.ac.jp, fukuda@eei.eng.osaka-u.ac.jp

Hirotsugu Kikuchi
Institute for Materials Chemistry and Engineering, Kyushu University, Kasuga, Fukuoka 816-8580, Japan

Funding: MEXT KAKENHI (JP23H02038, JP23H00303, JP18H03920 and JP25K23593), Grant-in-Aid for JSPS Fellows (JP23KJ1507, JP24KJ1622), JST ACT-X (JPMJAX24DE), and JST ASPIRE for Top Scientists / Top Teams (J251053026).

((**Abstract text.** Maximum length 200 words. Written in the present tense.))
Ferroelectric nematic liquid crystals (FNLCs) combine fluidity with spontaneous polarization, offering promising avenues for flexible electromechanical systems. Here, we demonstrate that mechano-electrical conversion in FNLCs can be enhanced by mechanically programming a robust macroscopic polarization alignment. Using hybrid liquid crystal cells composed of rigid glass and flexible substrates, we show that deformation in the ferroelectric nematic phase suppresses polarization domains and produces long-range ordered polarization alignment over millimeter-scale areas. This geometry-driven alignment originates from coupling between the FNLC's spontaneous splay deformation and the deformation-imposed cell geometry, and we further find that the selected polarization direction exhibits clear material dependence. Leveraging this deformation-enabled alignment, we develop an FNLC-based energy harvester that converts mechanical deformation into an output of approximately 1 V. These findings establish geometry-driven alignment as a practical design strategy for boosting FNLC mechano-electrical conversion while providing polarization control for soft electronic devices.

## 1. Introduction

Ferroelectric materials are indispensable in modern technology, underpinning applications ranging from non-volatile memories and piezoelectric actuators to electro-optic modulators and energy harvesting systems. The hallmark of these materials is the presence of spontaneous polarization that can be reoriented by external stimuli. While this phenomenon has long been associated with rigid crystal lattices, recent discoveries have revealed that fluid phases can also sustain long-range polar order.

The emergence of ferroelectric nematic liquid crystals (FNLCs) has introduced a new class of soft ferroelectrics that combine the fluidity and high symmetry of the nematic liquid crystal (NLC) phase with spontaneous polarization[1–3]. FNLCs exhibit spontaneous polarization of several μC $cm^{-2}$ near room temperature[3,4], together with giant dielectric anisotropy[3,5–9] and nonlinear optical phenomena[3,10–14]. These unique properties make FNLCs promising for next-generation soft electronics, photonics, and electromechanical devices, particularly in flexible and stretchable systems, where conventional solid-state ferroelectrics face limitations.

A defining strength of LCs lies in their controllable orientation. By appropriate surface treatments, the molecular director can be oriented in any desired direction, and its orientation can be reversibly switched under external fields such as electric or magnetic fields. This capability has been the cornerstone of the success of LC-based technologies, including flat-panel displays and phase-modulation devices[15].

This fundamental principle of orientation control applies equally to FNLCs. Their nonlinear optical responses, ferroelectric polarization, and electro-optic effects can be engineered and maximized through precise alignment control, prompting research into polarization control using rubbing and photoalignment. For instance, spray/bend patterns designed via photoalignment enable the patterning of polarization domains. This demonstrates that domain-controlled, high-quality alignment enables strong second-harmonic generation and geometric-phase control in FNLC optical elements, highlighting the device payoff of reliable alignment engineering[14,16,17]. Surface/anchoring effects are also becoming apparent. Rubbed or photoaligned interfaces with pretilt can impart polar anchoring that selects the sign of polarization[18–22]. This enables the formation of stable single domains in thin cells, laying the foundation for FNLC-based devices.

Beyond electro-optic control, FNLCs exhibit rich electro-mechanical and mechano-electric couplings that are highly relevant to both alignment control and innovative device concepts. Peter *et al.* demonstrated a viscous mechano-electric response: when FNLC is placed in a deformable container equipped with electrodes and mechanically actuated distinct electric current responses are generated due to both shape deformation and material flow[23]. Máthé *et al.* further reported linear electromechanical, converse piezoelectric, effects in FNLCs, with piezoelectric coupling constants exceeding 1 nC $N^{-1}$, comparable to or even surpassing those of strong solid piezoelectric materials[24]. Furthermore, they suggest the possibility of further performance improvements through alignment control. These results highlight the potential applications of tactile sensors, low-speed environmental mechanical energy harvesting, and soft actuators.

However, the key bottleneck for FNLCs is the difficulty of achieving uniform polarization alignment across macroscopic scales, which prevents realization of practical devices including mechano-electric devices. Depolarization fields and surface charge promote multidomain textures, particularly in thick cells, complicating deterministic alignment and reproducible device performance. This issue has been consistently reported from early FNLC research to recent analyses of equilibrium in twisted/splay domains[4,25–27]. Therefore, to achieve uniform orientation, it is necessary to introduce factors other than surface anchoring by the alignment film.

Here we show that mechano-electrical conversion in FNLCs can be significantly enhanced by mechanically programming a robust macroscopic polarization alignment. By coupling the spontaneous splay elasticity of the $N_F$ phase to deformation-imposed cell geometry, mechanical deformation drives the system into a uniform polarization alignment that is otherwise difficult to realize over large areas. Leveraging this geometry-driven alignment, we realize an FNLC energy harvester with an output of ~1 V. Moreover, we find that the selected polarization direction depends on the material, thereby providing fundamental insight that the intrinsic preferred direction of spontaneous splay deformation in FNLCs is material-specific. These results establish geometry-driven alignment as an effective design principle for enhancing mechano-electrical conversion while providing polarization control, paving the way for mechanically driven energy harvesters as well as other FNLC technologies in adaptive optics, soft robotics, and flexible electronics.

## 2. Results and Discussion

### 2.1. Polarization Alignment Induced by Mechanical Deformation

To investigate polarization alignment driven by mechanical deformation, we fabricated hybrid LC cells consisting of an ITO-coated rigid glass substrate paired with either a cover glass or a polyethylene naphthalate (PEN) film as flexible substrate. Three ferroelectric nematic liquid crystal (FNLC) materials were examined: two synthesized mixtures based on 1,3-dioxane derivatives (DIO Mixture 1 and DIO Mixture 2), and a commercial compound RM734. DIO Mixture 1 consists entirely of DIO-based molecules, whereas DIO Mixture 2 is a mixture of DIO-based molecules and Ester-based molecules. The molecular structures of their components are provided in **Figure 1**.

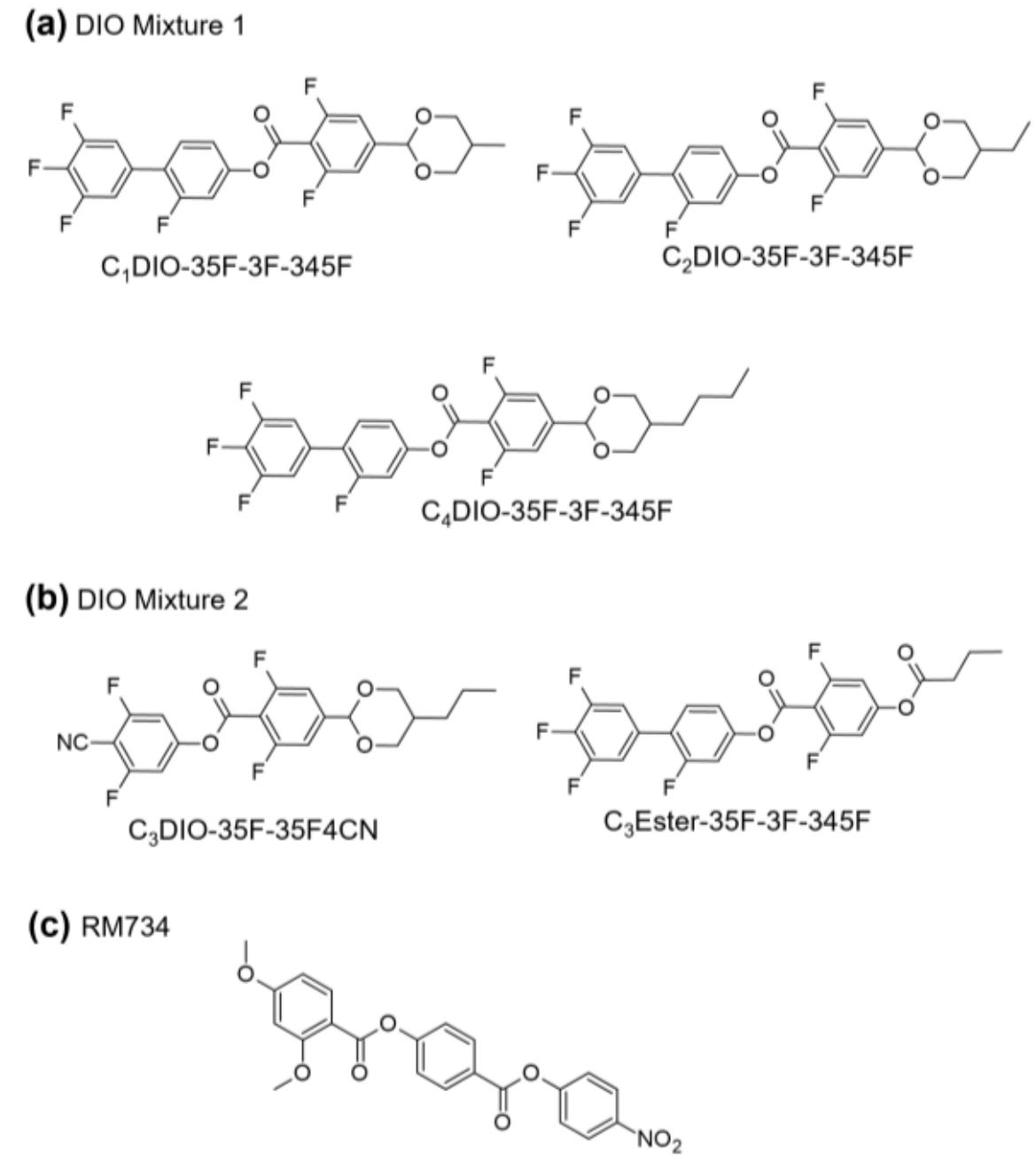

**Figure 1.** The molecular structures of FNLC materials: (a) DIO Mixture 1, (b) DIO Mixture 2, and (c) RM734.

We first examined the effect of deformation on the alignment state of FNLCs across different mesophases: the nematic (N) phase, an intermediate mesophase (M, including smectic $Z_A$), and the ferroelectric nematic ($N_F$) phase. For these experiments, a 10-μm-thick cell comprising a flexible cover glass and a rigid glass substrate was used to ensure identical surface conditions at both interfaces.

**Figure 2** shows representative polarized optical microscopy (POM) images of DIO Mixture 1 under relaxation and deformation. Mechanical strain was applied by pressing the flexible substrate with a tweezer tip, which appeared as a dark shadow in the upper left corner of the images. In the N phase, variations in color and brightness were observed as the cell thickness decreased upon compression; however, no significant textural changes were detected. This behavior suggests that strong surface anchoring at the glass interfaces stabilizes the alignment, thereby suppressing flow-induced reorientation. Upon release of the strain, the alignment fully recovered (**Figure 2a** and **Movie S1**).

In the M phase, deformation led to the disappearance of the multidomain structure near the center of the deformed region, where a radial pattern emerged, indicating partial radial alignment of the FNLC molecules. This alignment, however, was confined to a limited area, and the surrounding multidomain texture remained largely unaffected (**Figure 2b** and **Movie S2**).

A markedly different response was observed in the $N_F$ phase. Here, deformation erased the preexisting domain boundaries entirely, giving rise to concentric circular patterns reflecting thickness variations and distinct radial dark lines corresponding to radial polarization alignment. Notably, this mechanically induced alignment propagated over long distances of several millimeters and persisted for extended durations, even under continuous deformation lasting several seconds (**Figure 2c** and **Movie S3**). Similar phenomena were consistently observed in DIO Mixture 2 and RM734 **(Figures S1 and Figures S2, respectively)**.

These findings demonstrate that polarization alignment in the $N_F$ phase is not governed by flow or surface alignment layers but is instead dictated by confinement within the deformed geometry. Since the $N_F$ phase is known to exhibit pronounced intrinsic splay elasticity, we attribute the observed alignment to the effective coupling of this elasticity with the cell geometry, enabling robust and reversible mechanical control of the polarization direction.

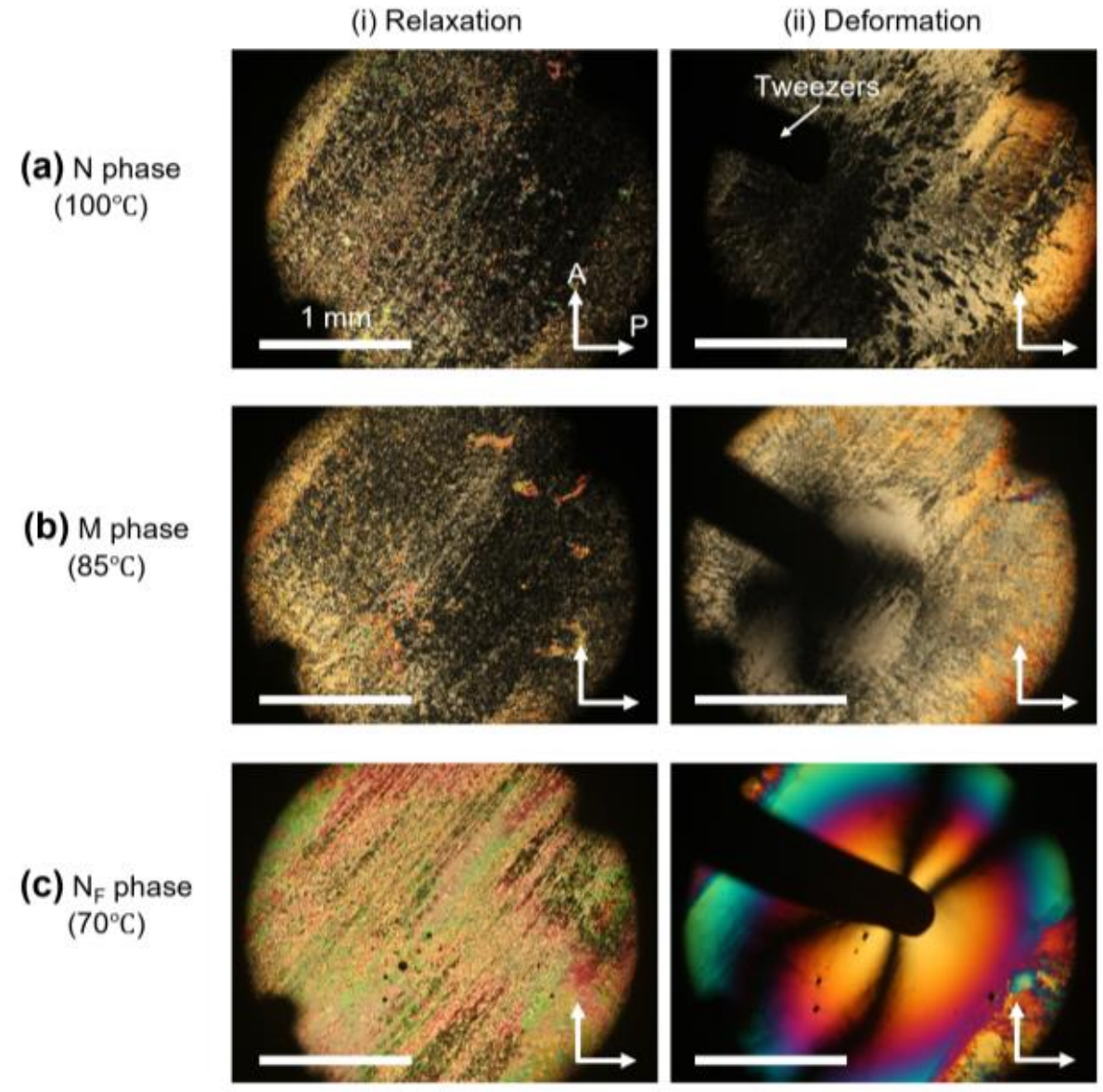

**Figure 2.** POM images of DIO Mixture 1 under (i) relaxation and (ii) mechanical deformation in (a) N, (b) M, and (c) $N_F$ phases. In the $N_F$ phase, deformation erased domain boundaries and induced long-range radial polarization alignment.

### 2.2. Evaluation of Polarization Direction under Mechanical Deformation

To evaluate the polarization direction induced by mechanical deformation, we measured the voltage signals generated during the application and release of stress in the system illustrated in **Figure 3a**. The LC cell was directly connected to an oscilloscope, and the current was monitored through the internal impedance of 1 MΩ. Mechanical deformation was applied by pressing the tweezer tip against the ground electrode side of the flexible substrate, inducing a tapered shape where the thickness on the ground side decreased and the thickness on the opposing electrode side increased. The deformation was maintained for approximately 1 second and then released.

**Figure 3b-d** summarizes the voltage waveforms obtained in each phase for all FNLC materials. In both the N and M phases, no significant voltage signals were detected. Although flexoelectric polarization should in principle occur in the N phase, its magnitude is expected to be extremely small. A similar situation applies to the M phase. Despite the presence of short-range order, the absence of long-range order leads to mutual cancellation of local polarizations, even if molecules are transiently aligned near the deformed region.

In striking contrast, in the $N_F$ phase two distinct voltage peaks were observed, corresponding respectively to the application and release of deformation. Notably, the polarity of the voltage signals differed among the FNLC materials. Whereas DIO Mixture 1 exhibited a negative voltage signal during deformation, DIO Mixture 2 and RM734 produced positive signals. These waveforms indicate that electrons flow either toward or away from the deformed region during strain, depending on the material. The inferred polarization directions are schematically illustrated in **Figure 3b-d(iv)**. In DIO Mixture 1 the polarization points toward the tapered end, whereas in DIO Mixture 2 and RM734 it points toward the broader end. This finding clearly demonstrates that the spontaneous splay deformation direction varies depending on the specific FNLC material.

It is worth noting that the observed differences cannot be simply attributed to molecular shape. RM734 possesses a wedge-like structure with its wider end oriented along the polarization direction, whereas the DIO-based molecules are nearly rod-like, with only a slight taper. In conventional flexoelectric effects, polarization is typically induced through coupling between molecular geometry and director curvature. However, the present results do not conform to such a straightforward picture. This suggests that the spontaneous splay deformation in FNLCs cannot be explained solely by molecular shape. The major difference between DIO Mixture 1 and DIO Mixture 2 is that while the constituent molecules in DIO Mixture 1 are all DIO-based, DIO Mixture 2 is a mixture of DIO-based molecules and Ester-based molecules. Consequently, in DIO Mixture 2, intermolecular packing is weaker, and the steric effect of the alkyl chains is likely to be more pronounced, and the polarization direction may have pointed towards the broad end. However, the microscopic mechanism that determines the sign of spontaneous splay in FNLCs remains unresolved. Although molecular geometry, dipolar interactions, and local molecular packing are all expected to contribute, their relative roles and collective effects are not yet fully understood. Clarifying how these factors cooperatively dictate the direction of polarization is an open challenge in FNLC research, and further systematic investigation will be required to establish a comprehensive microscopic model.

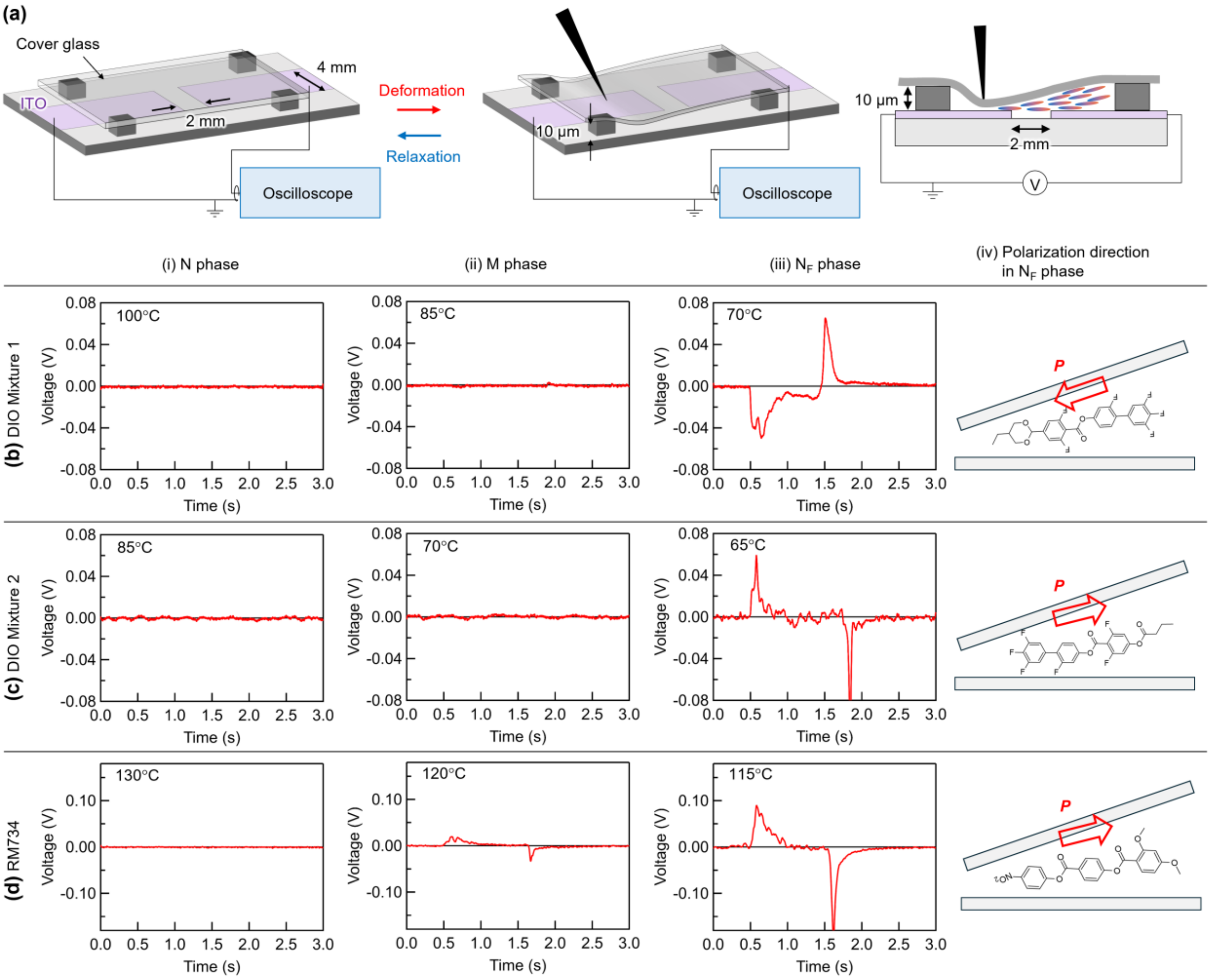

**Figure 3.** (a) Schematic of the experimental setup for voltage measurement under mechanical deformation of FNLC cells. Mechanical stress was applied to the flexible side of the cell with a tweezer tip for ~1 s and then released. This deformation of the flexible substrate induces a tapered shape where the cell thickness on the grounding side decreases and the thickness on the opposing electrode side increases. (b–d) Voltage waveforms of (b) DIO Mixture 1, (c) DIO Mixture 2, and (d) RM734 recorded in (i) N, (ii) M, and (iii) $N_F$ phases. Significant signals were only observed in the $N_F$ phase, showing two peaks corresponding to deformation

and release. (iv) Schematic illustration of the inferred polarization direction in the $N_F$ phase for each FNLC material.

### 2.3. Effect of Substrate Alignment on Mechanically Induced Polarization

FNLCs are known to be highly sensitive to substrate properties, such as fluorophilicity and pretilt, which can strongly affect their alignment. To examine how surface anchoring influences mechanically induced polarization, we investigated DIO Mixture 1 in two types of cells: planar-degenerate alignment and vertical alignment.

In the relaxed state of the planar-degenerate cell (**Figure 4a**), a striped multidomain texture was observed due to depolarization. Upon mechanical deformation, the stripe pattern persisted, but alternating dark and bright regions appeared. This behavior can be interpreted as follows: when the intrinsic polarization direction within a domain is nearly parallel to the deformation-induced alignment force, the molecules align uniformly, resulting in a dark contrast; when the directions are misaligned, the texture remains bright. These observations suggest that although a deformation-induced aligning torque is present, it is much weaker than the surface anchoring imposed by the alignment layer. Consistently, only negligible voltage signals were detected, indicating that the interfacial alignment state was hardly altered.

**Figure 4b** shows POM images of the vertical alignment cell under deformation. A degree of polarization alignment was observed immediately upon pressing. However, this order decayed rapidly, and domain formation proceeded shortly thereafter. In fact, the onset of domain development is already evident within the image. No discernible electrical signals were detected during either deformation or relaxation. Unlike the large voltage signals observed in ITO/glass cells shown in **Figure 3**, the absence of a signal in vertical alignment cells can be attributed to the lack of significant out-of-plane polarization changes. Although in-plane rotations may occur, the vertical anchoring prevents variations in the polarization rise direction, resulting in little change in the surface charge stored at the ITO electrodes. Additionally, the absence of detectable voltage signals may also be influenced by charge accumulation at the LC–alignment layer interface. The formation of an electric double layer could destabilize the bulk polarization alignment through electrostatic interactions, leading to rapid depolarization. Further systematic studies, for example under controlled ionic conditions or with different alignment layers, will be required to verify this mechanism.

Taken together, these results demonstrate that in both planar and vertical alignment cells, the presence of alignment layers suppresses mechanically induced polarization alignment. This indicates that alignment force of flexoelectric coupling is intrinsically weak, and highlights the critical importance of carefully considering interfacial anchoring effects in the design of FNLC devices.

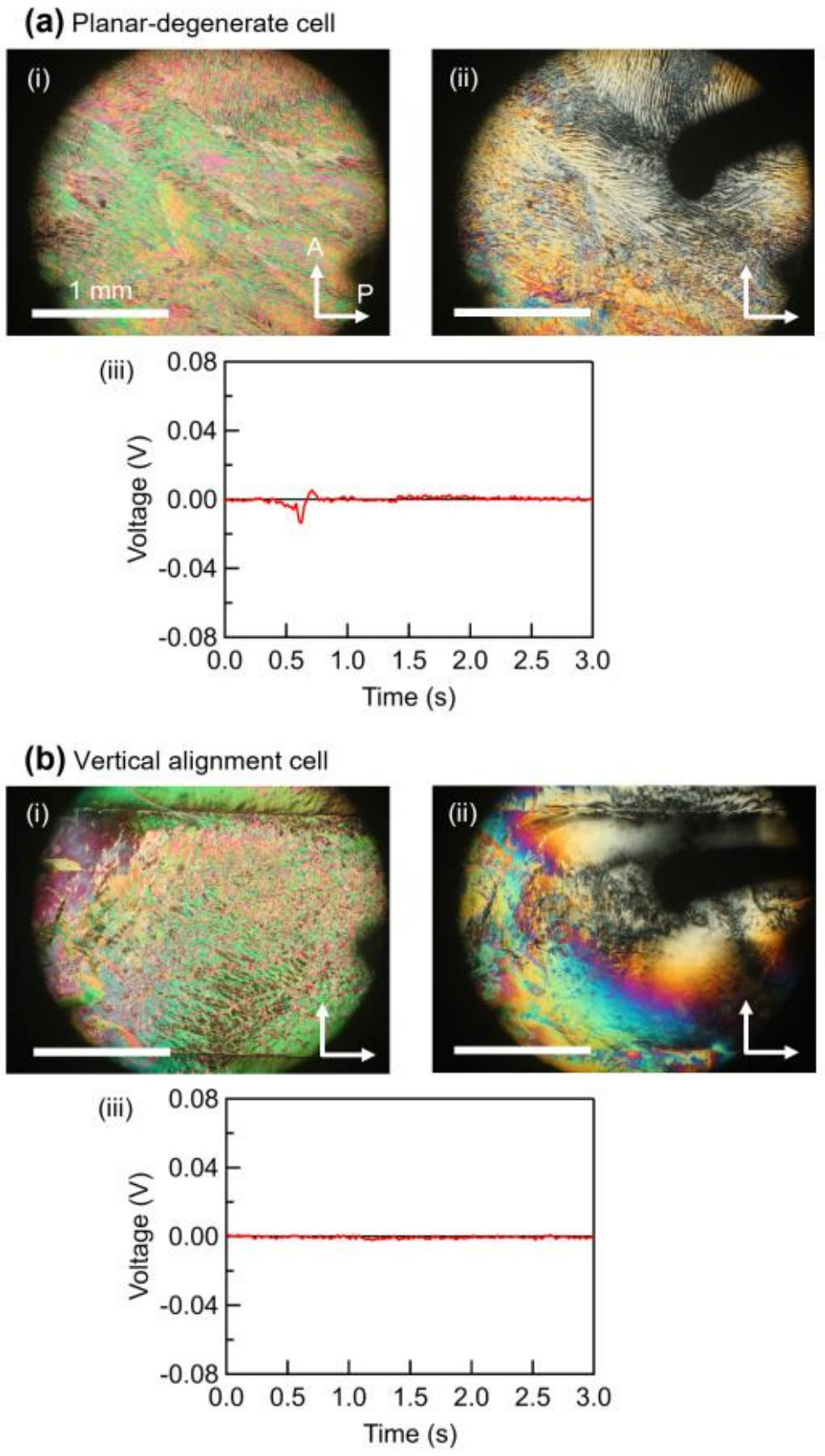

**Figure 4.** Effect of surface alignment treatment on the mechanically induced polarization alignment of DIO Mixture 1 in (a) planar degenerate cells and (b) vertical alignment cells: (i) relaxed state, (ii) under deformation, (iii) voltage waveform corresponding to deformation.

### 2.4. Strain-Amplitude Dependence of Polarization

To investigate the relationship between strain amplitude and polarization, we employed a 180-μm-thick cell composed of a polyethylene naphthalate (PEN) substrate and glass, which allows for large mechanical deformation. Here, the PEN substrate was supported at one end, forming a cantilever geometry. Using the configuration shown in **Figure 5a**, the PEN substrate was displaced by an arbitrary amount $dz$ with a micrometer screw and then released. The accumulated charge was calculated from the current flowing into the oscilloscope. As shown in **Figure 5b**, the accumulated charge increased monotonically with small deformation amplitudes, remaining at a low level. When the deformation exceeded approximately 70 μm, the charge increased abruptly and eventually saturated near a deformation amplitude of 100 μm. This threshold-like behavior suggests that a critical mechanical strain is required for flexoelectric coupling to overcome the energetic advantage of domain cancellation and to stabilize a uniformly polarized state. Below this threshold, FNLCs remain in the energetically favorable multidomain configuration, resulting in negligible macroscopic polarization. It should be emphasized, however, that this behavior arises from a combination of several factors, including cell thickness and interfacial anchoring. As demonstrated in **Figure 3**, surface anchoring strongly influences polarization alignment. In the present case, PEN substrates were used to achieve large deformation amplitudes, but the strain–polarization relationship may not necessarily be identical to that observed in glass-based cells. Moreover, the role of electric double layers at the substrate interface and depolarization in the bulk suggests that the results may also strongly depend on the cell thickness. While conventional flexoelectric effects are typically discussed in terms of combination of splay and bend deformations[28], it remains unclear whether the same framework can be directly applied to flexoelectric coupling in FNLCs. We also evaluated the anchoring effect associated with flexoelectric coupling by applying a triangular-wave voltage under a deformation of 180 μm and measuring the polarization reversal current (**Figure S3**). Although the symmetry of the reversal process was slightly broken by flexoelectric coupling, only minor changes were observed. This weak effect is likely attributable not only to the intrinsically weak anchoring strength but also to the high viscosity and elasticity of FNLCs, which result in slow relaxation toward a new equilibrium state.

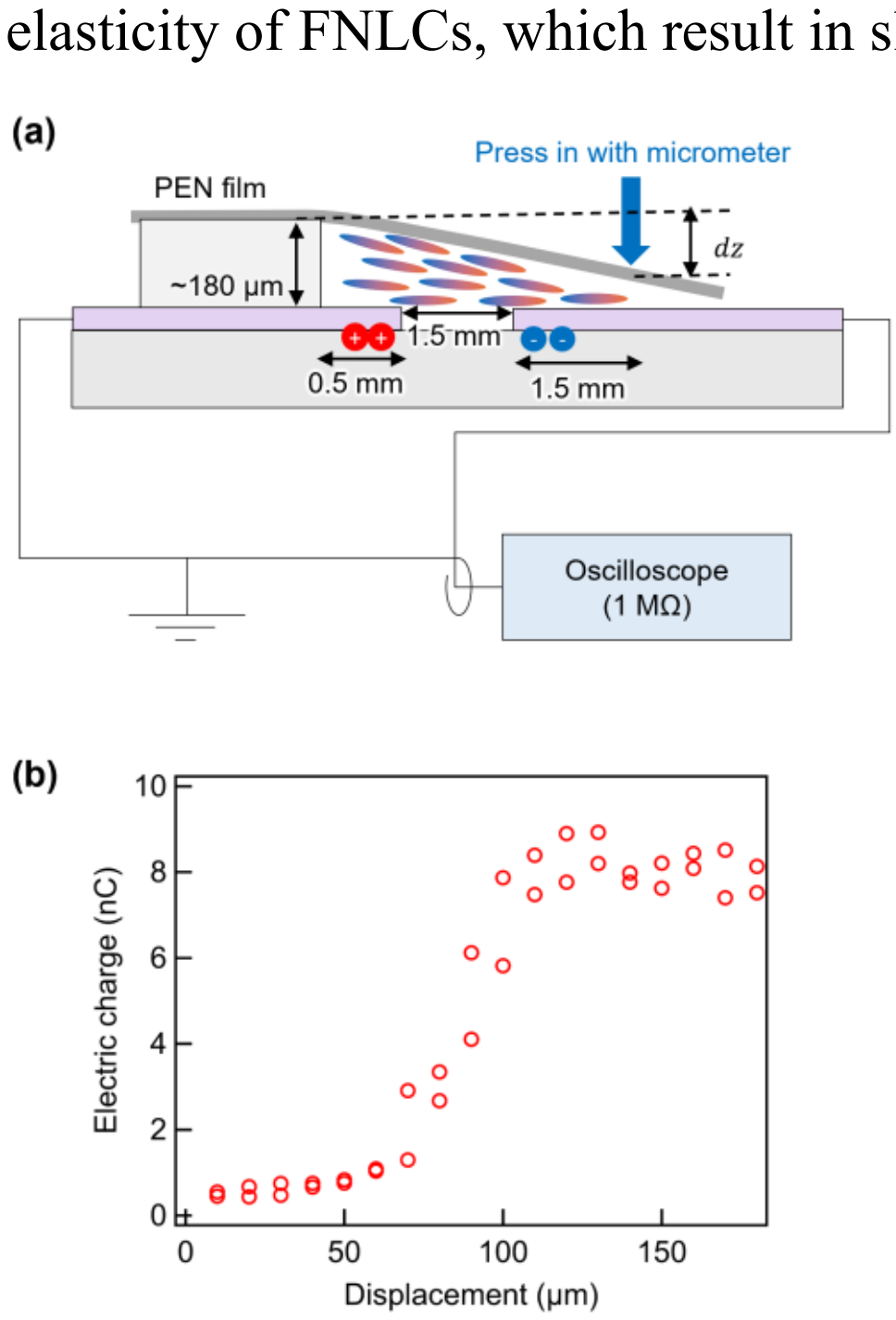

**Figure 5.** (a) Experimental setup for strain–polarization measurements using a 180-μm-thick FNLC cell with a PEN substrate. The PEN substrate was supported at one end, forming a cantilever geometry as illustrated. (b) Accumulated charge as a function of displacement.

## 2.5. Energy Harvesting via Polarization Reversal

Finally, we explored the application of mechanically induced polarization switching for energy harvesting. We constructed FNLC cells with electrodes positioned on both sides of a mechanically deformable region, as shown in **Figure 6a**. During experiments, we repeatedly applied and released mechanical deformation at the center of the FNLC cell while monitoring the voltage output in real time. Representative voltage waveforms are shown in **Figure 6b**. Upon deformation, a positive voltage peak was consistently observed, whereas the release of deformation induced a negative peak. These alternating voltage signals indicate that the process involves genuine polarization-depolarization cycles within the FNLC layer rather than a mere transient displacement current.

Quantitatively, under a 1 Hz operation (one deformation–release cycle per second) and connected to the oscilloscope impedance $R_L = 1$ MΩ, the areal average power density $\bar{P}_{areal}$ was $\sim 0.02$ μW $\text{cm}^{-2}$. This value was obtained by integrating the measured waveform over one cycle via $E_{cycle} = \int V(t)^2/R_L\, dt$ and converting to average power per unit area, $\bar{P}_{areal} = E_{cycle}/T$ (with $T = 1$ s). The signals were stable and reproducible over multiple deformation cycles, demonstrating good reversibility of the energy conversion mechanism.

Because the present device employs an in-plane electrode configuration, most of the spontaneous polarization in a uniformly aligned FNLC does not contact the electrodes, and the effective coupling to the bulk polarization is inherently limited. A simple upper-bound estimate indicates that, if the device were re-engineered to extract the full out-of-plane polarization (e.g., by adopting an out-of-plane electrode architecture), the areal average power density of ~8 μW $\text{cm}^{-2}$ at 1 Hz. While this idealized value neglects losses and structural limitations which need to induce spray deformation, it delineates a realistic target for device optimization. Importantly, this estimated upper-limit value is comparable to the performance of conventional piezoelectric vibration harvesters, suggesting that FNLC-based devices could achieve energy conversion efficiencies on par with established rigid piezoelectric counterparts despite being extremely thin, soft, and stretchable.

These considerations underscore clear pathways for performance improvement. Future strategies, such as out-of-plane electrode designs, structure design to induce splay deformation and anchoring conditions, should enhance the electromechanical coupling and raise the areal energy and power densities toward the upper-bound regime. Taken together, our results establish FNLCs as promising candidates for soft, flexible, and mechanically driven energy harvesters, offering a mechanistically distinct route to transduce deformation into electrical power via polarization dynamics.

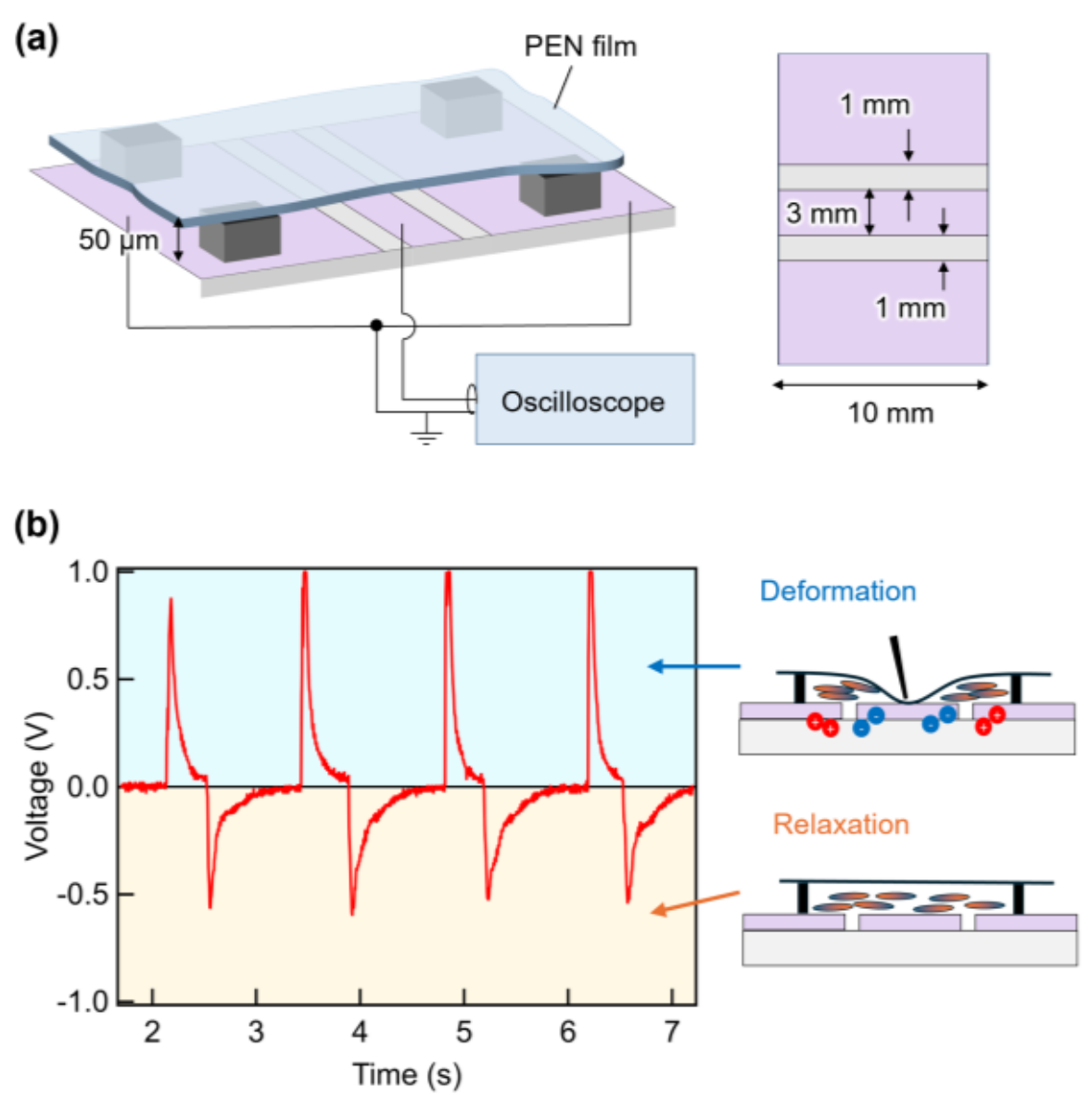

**Figure 6.** (a) Schematic of the FNLC energy-harvesting device with a PEN substrate.
(b) Voltage output showing reproducible positive and negative peaks upon cyclic deformation and relaxation, demonstrating reversible polarization switching.

## 3. Conclusion

In this work, we demonstrated that mechano-electrical conversion in FNLCs can be effectively enhanced by mechanically programming a robust macroscopic polarization alignment. Specifically, mechanical deformation suppresses polarization domains and drives the $N_F$ phase toward long-range alignment through coupling between spontaneous splay elasticity and deformation-induced confinement. Building on this deformation-enabled alignment, we realized a proof-of-concept FNLC energy harvester that generated reproducible alternating voltage signals and measurable power densities under cyclic strain. Strain–response analysis further revealed a threshold behavior that emerges from the competition between domain stabilization and deformation-driven alignment, providing a practical guideline for device operation. Notably, the selected polarization direction was material-dependent, suggesting that collective molecular interactions, rather than simple molecular geometry, govern the preferred sense of spontaneous splay. Overall, these results establish deformation-based polarization control as a practical design principle for boosting FNLC mechano-electrical conversion while overcoming the long-standing challenge of macroscopic polarization alignment, opening pathways toward flexible sensors, actuators, and energy-harvesting systems.

## 4. Methods

### 4.1 Materials

Three ferroelectric nematic liquid crystal (FNLC) materials were employed in this study: Two synthesized mixtures based on 1,3-dioxane-derived mesogenic units (designated as DIO Mixture 1 and DIO Mixture 2), and commercially available 4-((4-Nitrophenoxy)carbonyl)phenyl 2,4-dimethoxybenzoate (RM734, CHEMFISH). Here, DIO Mixture 1 is a mixture of DIO analogues with terminal alkyl chain lengths of 1, 2, and 4, and DIO Mixture 2 is a mixture of DIO analog with a terminal alkyl chain length of 3 and a cyano group on the opposite side, and Ester analog where the 1,3-dioxane units converted to ester units. The molecular structures of the constituent molecules in each FNLC material are illustrated in **Figure 1**. The phase transition temperatures for each material were determined by POM observation and are summarized in **Figure S4**.

### 4.2 Cell fablication

To enable mechanical deformation experiments, LC cells were fabricated using hybrid substrates composed of a flexible and a rigid component. The rigid substrate consisted of indium tin oxide (ITO)-coated glass, selected for its transparency and electrical conductivity to facilitate both optical observation and measurement of generated voltages. The flexible substrates used were either cover glass (thickness ~150 μm) or polyethylene naphthalate (PEN) films (thickness ~150 μm). Cover glass was employed to maintain consistent surface properties with the rigid glass substrate, while PEN films were selected in cases where larger deformation amplitudes were required due to their superior mechanical flexibility. Surface treatments of substrates varied across three configurations; untreated (cleaned only, no alignment layer), coated with a polyimide-based planar alignment layer (AL1254, JSR), and coated with a polyimide-based vertical alignment layer (JALS-2024-R2, JSR). For cells constructed from cover glass and rigid glass substrates, the cell gap was set to 10 μm using silica spacers. For cells using PEN film and glass substrates, the cell gap was set 50 μm or 150 μm using PET film spacers.

### 4.3 Optical and electrical characterization

All LC cells were placed on a temperature-controlled stage (T95-PE, Linkam) to enable precise regulation of sample temperature during experiments, ensuring accurate investigation of different LC phases. Mechanical deformation was applied manually using fine tweezers while simultaneously observing the LC textures under a polarized optical microscope (Eclipse LV100 POL, Nikon).

To characterize polarization-related electrical signals, the electrodes of the LC cells were connected directly to a digital storage oscilloscope (TDS2024B, Tektronix) with an input impedance of 1 MΩ. Voltage waveforms generated during mechanical deformation and relaxation were monitored in real time, enabling quantitative assessment of the FNLC materials' electromechanical response.

**Acknowledgements**

Partial components of the liquid crystal mixture were provided by JNC Petrochemical Corporation. This work was partly supported by MEXT KAKENHI (JP23H02038, JP23H00303, JP18H03920 and JP25K23593), Grant-in-Aid for JSPS Fellows (JP23KJ1507, JP24KJ1622), JST ACT-X (JPMJAX24DE), and JST ASPIRE for Top Scientists / Top Teams (J251053026).

**Data Availability Statement**

All data supporting the findings of this study are available within the Article and its Supplementary Information.

# Supporting Information

**Geometry-Driven Polarization Control in Ferroelectric Nematic Liquid Crystals**

*Kazuma Nakajima*, Hirokazu Kamifuji, Hirotsugu Kikuchi, Kenjiro Fukuda* and Masanori Ozaki*

Kazuma Nakajima, Hirokazu Kamifuji, Kenjiro Fukuda and Masanori Ozaki
Department, University, City, Country
Division of Electrical, Electronic, Informational Engineering, Graduate School of Engineering, The University of Osaka, 2-1 Yamada-oka, Suita, Osaka 565-0871, Japan
E-mail: nakajima.kazuma@eei.eng.osaka-u.ac.jp, fukuda@eei.eng.osaka-u.ac.jp

Hirotsugu Kikuchi
Institute for Materials Chemistry and Engineering, Kyushu University, Kasuga, Fukuoka 816-8580, Japan

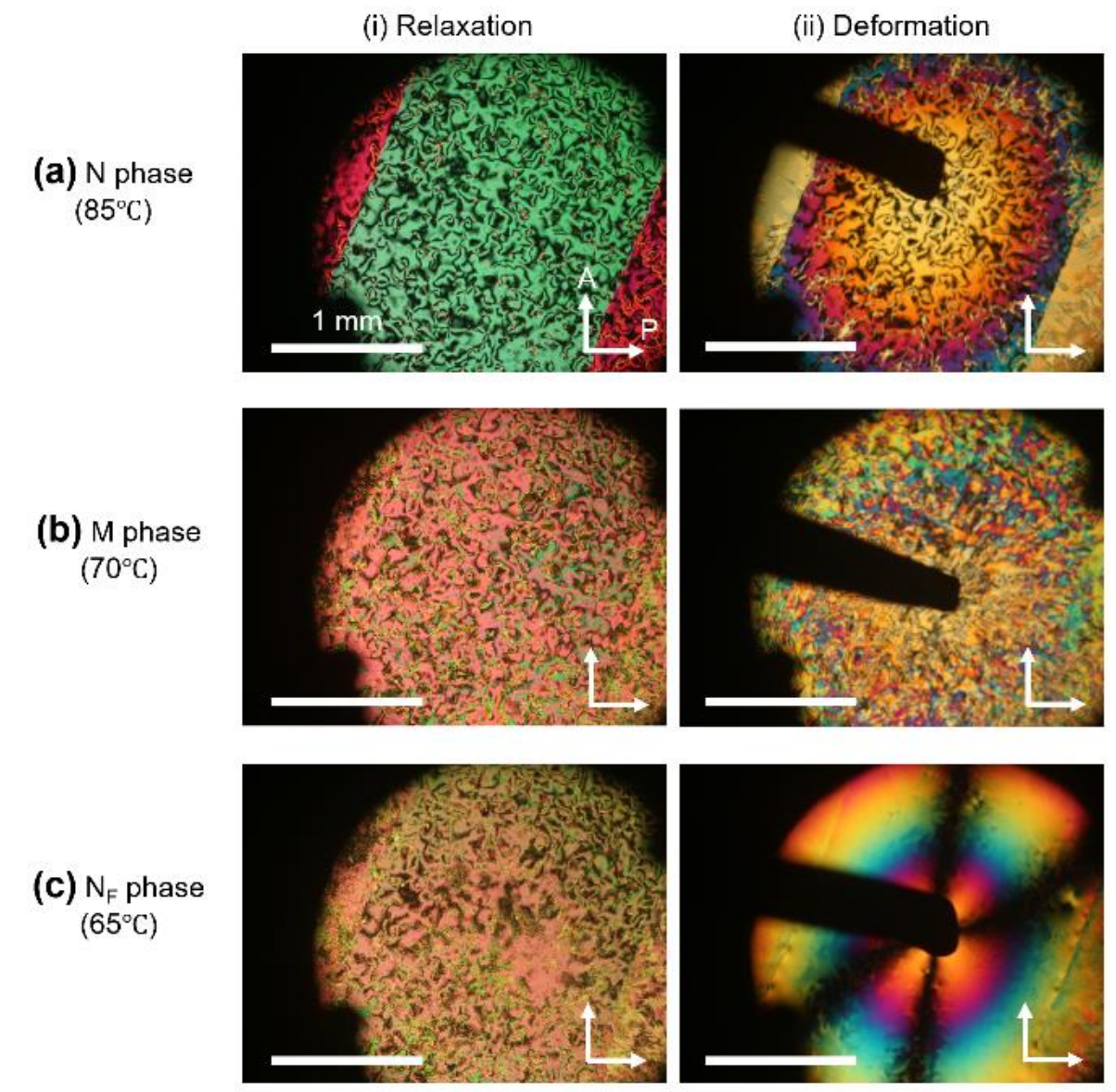

**Figure S1.** Polarized optical microscopy (POM) images of DIO Mixture 2 under (i) relaxation and (ii) mechanical deformation in (a) N (85°C), (b) M (70°C), and (c) $N_F$ phases (65°C).

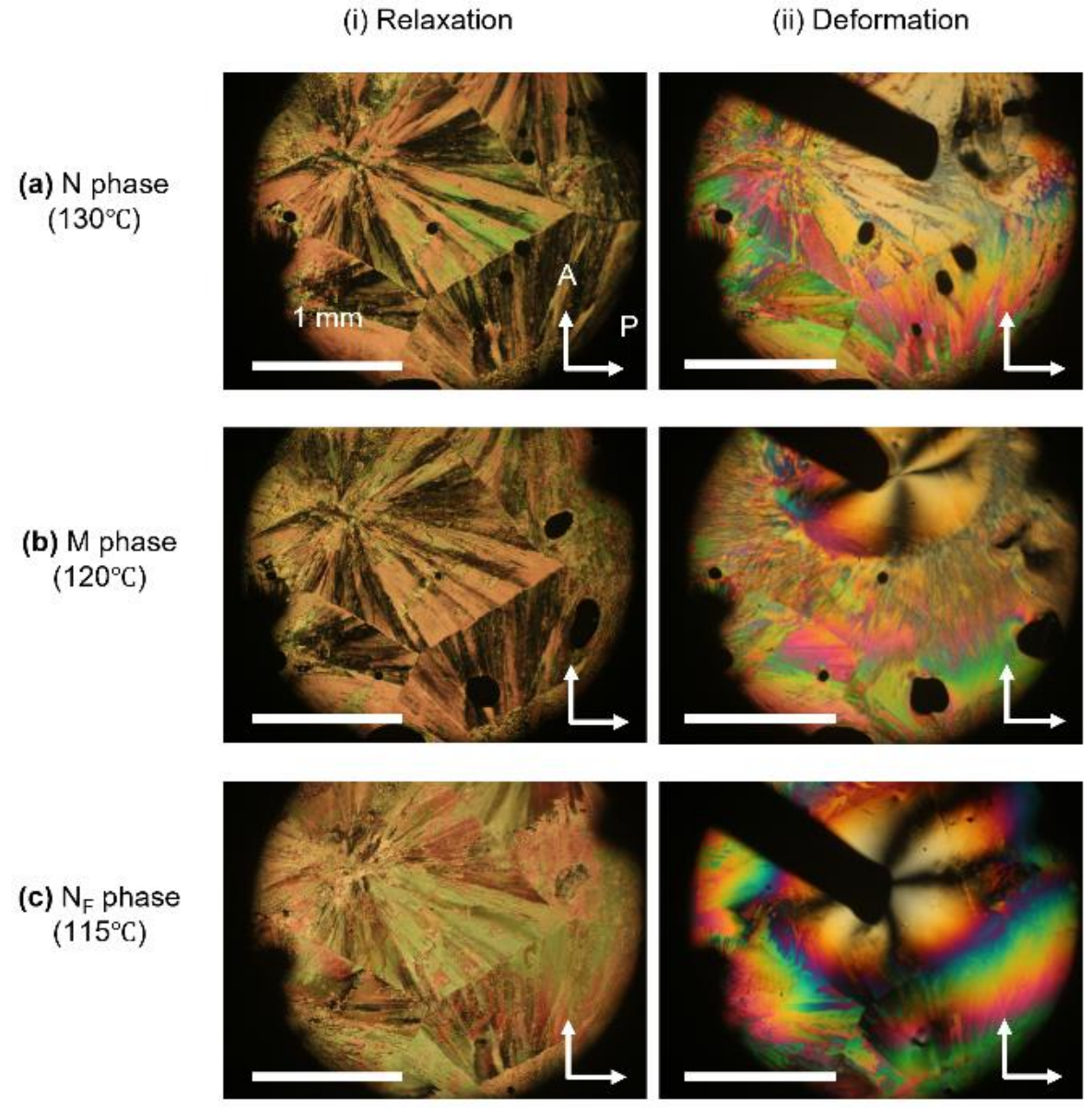

**Figure S2.** POM images of RM734 under (i) relaxation and (ii) mechanical deformation in (a) N (130°C), (b) M (120°C), and (c) $N_F$ phases (115°C).

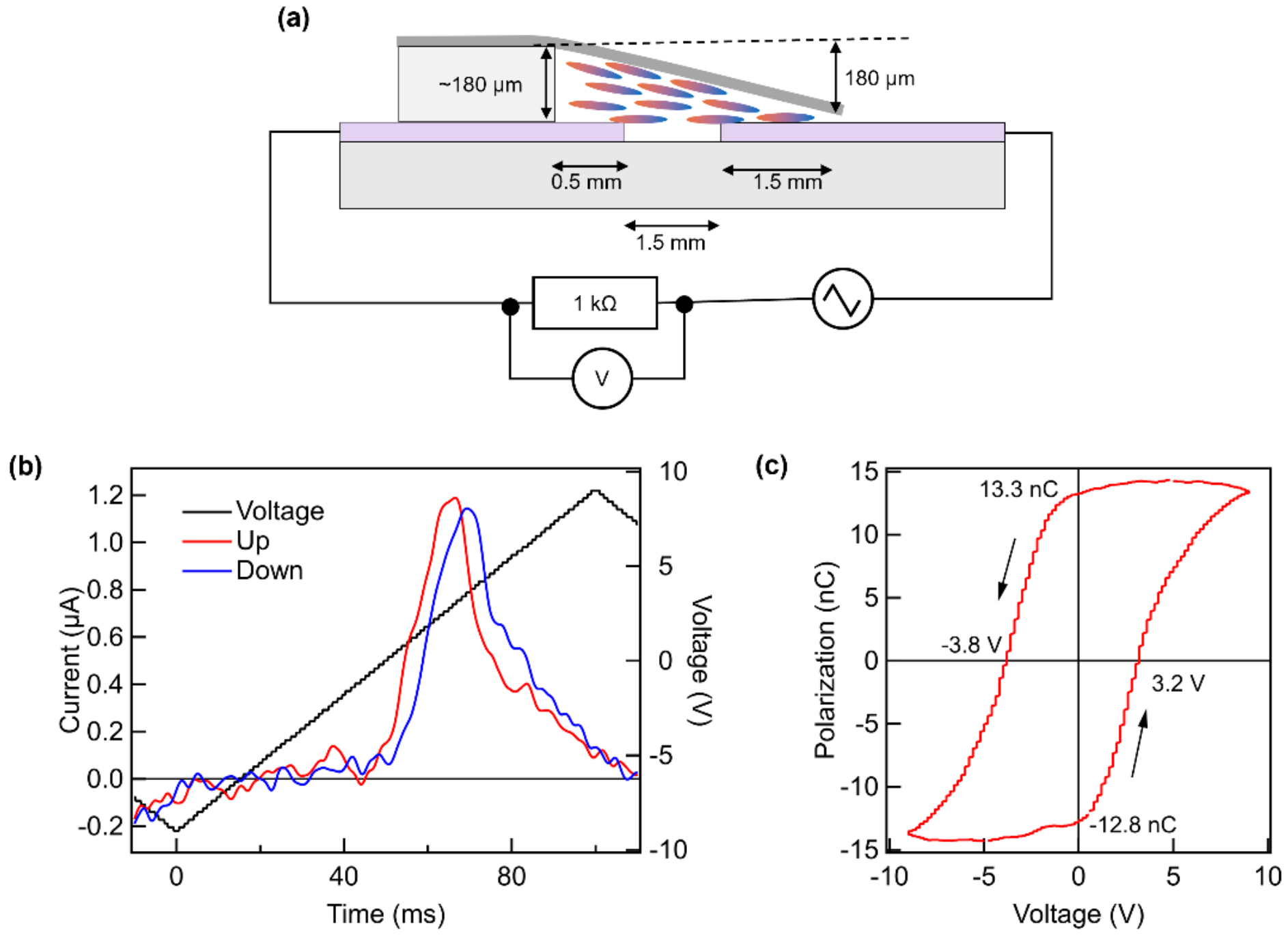

**Figure S3.** (a) Schematic illustration of the polarization reversal current waveform measurement under mechanical deformation. (b) Polarization reversal current waveform observed when a triangular voltage of 5 Hz and 10 $V_p$ was applied. Up corresponds to the process in which the voltage on the thicker side of the cell increases, and Down corresponds to the process in which the voltage decreases, with the current flowing through a 1 kΩ resistor. (c) Hysteresis loop calculated from the polarization reversal current waveform. A slight asymmetry was observed due to the alignment force induced by flexoelectric coupling.

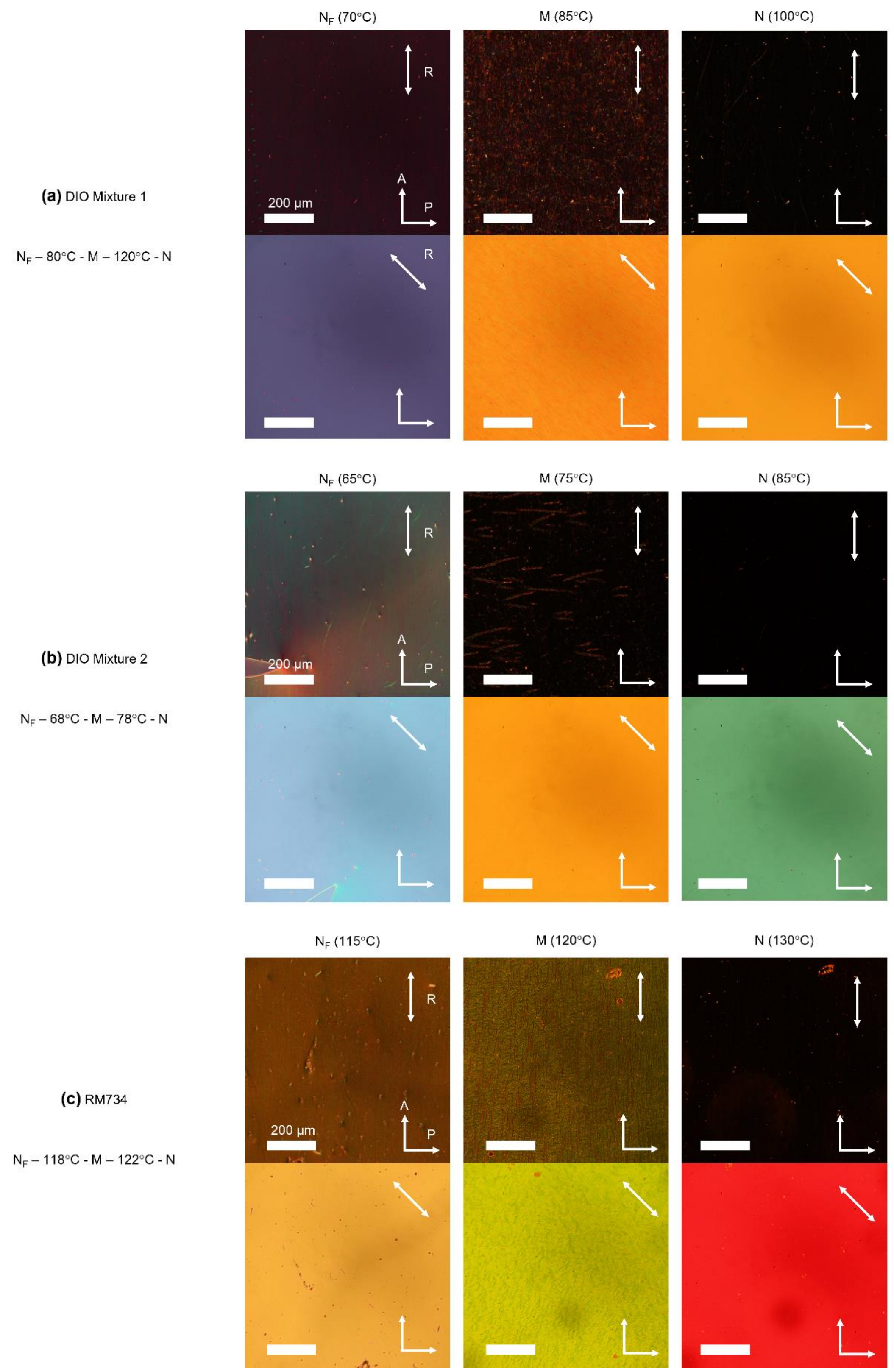

**Figure S4.** Phase transition temperatures and typical POM images of each phase for (a) DIO Mixture 1, (b) DIO Mixture 2, and (c) RM734. Here, a rubbing cell with a thickness of 6 μm was used, where the rubbing directions were anti-parallel.